\documentclass {article}
\usepackage{graphicx}
\usepackage{float}
\usepackage{url}
\usepackage{braket}
\usepackage{setspace}
\usepackage{amsmath,amsthm,amssymb,bm}
\newcommand{\gtsim}{\protect\raisebox{-0.5ex}{$\:\stackrel{\textstyle >}{\sim}\:$}} 

\def\vec#1{\mbox{\boldmath $#1$}}

\begin{document}

\topmargin =-20mm
\title{Testing Bell's Inequality using Charmonium Decays}
\author{Shion Chen$^{1}$\thanks{chen@icepp.s.u-tokyo.ac.jp},  Y\={u}ki Nakaguchi$^{1, 2}$ and Sachio Komamiya$^1$ \\ \\
\fontsize{10.0pt}{0pt}\selectfont ${}^1$ Department of Physics, Graduate School of Science, The University of Tokyo\\
\fontsize{10.0pt}{0pt}\selectfont 7-3-1, Hongo, Bunkyo-ku, 113-0033, Tokyo, Japan \\
\fontsize{10.0pt}{0pt}\selectfont ${}^2$  Institute for the Physics and Mathematics of the Universe, The University of Tokyo \\
\fontsize{10.0pt}{0pt}\selectfont 5-1-5, Kashiwano-ha, 277-8583, Chiba, Japan}
\date{}

\setlength{\abovedisplayskip}{4pt} 
\setlength{\belowdisplayskip}{8pt}

\maketitle

\begin{center}
\textbf{Abstract} \\
\end{center}
This paper discusses the feasibility of testing Bell's inequality with the charmonium decays $\mathrm{\eta_c \rightarrow \Lambda\overline{\Lambda}}$, $\mathrm{\chi_{c0} \rightarrow \Lambda\overline{\Lambda}}$ and  $\mathrm{J/\psi \rightarrow \Lambda\overline{\Lambda}}$. We develop a new formulation of the Bell's inequality represented with particles' orientations which can be measured straightforwardly in experiment, and seek if it violates in these decay processes. It is illustrated that the $\eta_c$ channel and the $\chi_{c0}$ channel maximally violate Bell's inequality, while the J/$\psi$ channel gives no inconsistency with it. Simulations dealing with experimental aspects are also implemented. The expected statistical fluctuations and achievable significance are estimated as a function of sample size.

\topmargin =-13mm

\section{Introduction}
With no doubt, Bell's inequality (BI) has played a significant role in modern physics. It establishes a clear discrimination between quantum mechanical and classical view of nature. Since when it was first shown to be violated in a photons system by optical experiments \cite{Alain}, what if with other type of particles or systems have long been people's interest. Cases involving massive particles are in particular curiosity, since generally a massive particle displays more classical properties. At present, some of the experiments have been implemented dealing with proton pairs \cite{Sakai} or  $\mathrm{K^0 \overline{K^0}}$ and $\mathrm{B^0\overline{B^0}}$ oscillations \cite{KK}\cite{BB}, however the variety of the experiments is still limited. (See also the review by Y. B. Ding {\it et.al.} \cite{review}) \\
In this context, charmonium decays $\mathrm{ c\overline{c} \rightarrow \Lambda \overline{\Lambda}}$ have been anticipated to be one of the testing channel of BI \cite{Ton1981}\cite{Ton1986} in a highly exotic system: it involves high energy, massive, unstable fermions, experiencing parity-violating weak interaction. In particular, $\eta_c \rightarrow \Lambda\overline{\Lambda}$ and $\chi_{c0} \rightarrow \Lambda\overline{\Lambda}$ are exact realizations of Bohm's type of EPR experiment \cite{Borm} in that these involve a spinless particle in the initial state decaying into two spin one half fermions with opposite spins. In quantum mechanics (QM) this is known as an entangled state, leading to a strong correlation between the two fermions, that violates BI. $J/\psi \rightarrow \Lambda\overline{\Lambda}$ is similar, and is thought to be the most promising channel because of the abundant statistics. However this channel has the significant difference of having a spin 1 initial state. This allows involvement of relative orbital angular momentum between the generated hyperons $\Lambda$ and $\overline{\Lambda}$, which disturbs the entanglement and weakens the particle correlation. It has not been clear if this weakened correlation is nonetheless sufficient to violate BI (We discuss and give a conclusion to this in chapter 4).



Practically, this test seems to include a complication that we have no means to directly measure the spin of decaying hyperons. For an indirect approach instead, it has been suggested by N. A. Tornqvist that the association between $\Lambda$ ($\overline{\Lambda}$) spin and the decay distribution of its weak decay $\mathrm{\Lambda \rightarrow \mathrm{p} \pi^- }$ ($\mathrm{\overline{\Lambda} \rightarrow \overline{\mathrm{p}}\pi^+}$) is useful for inferring the hyperon spins within the classical picture \cite{Ton1981}.
The distribution of this decay product proton (anti-proton) is known to be
\begin{equation}
\frac{d\Gamma_{\Lambda}}{d\Omega_p} \propto 1 +  \alpha_{\Lambda} \vec{n} \cdot \vec{s}  \mbox{\phantom{MMMMMM}}
\frac{d\Gamma_{\bar{\Lambda}}}{d\Omega_{\bar{p}}} \propto 1 -  \alpha_{\bar{\Lambda}} \vec{n'} \cdot \vec{s'}
\label{eq1}
\end{equation}
in the $\Lambda$ ($\overline{\Lambda}$) rest frame \cite{Lambda}. $\vec{n}$ (\vec{n'}) is the unit vector of the outgoing proton (anti-proton) direction and $\vec{s}$ ($\vec{s'}$) the polarization vector of the $\Lambda$ ($\overline{\Lambda}$) as shown in Fig.2, which are all defined in the $\Lambda$ ($\overline{\Lambda}$) rest frame. 
$\alpha_{\Lambda}$ and $\alpha_{\bar{\Lambda}}$ are the decay parameters with experimental values of $\alpha_{\Lambda} = 0.642 \pm 0.013$ \cite{PDG} and $\alpha_{\Lambda} = \alpha_{\bar{\Lambda}}$ if ignoring the small effect of CP violation. Due to the parity-violating nature of the weak interaction, the outgoing proton (anti-proton) prefers to fly along (against) the polarization of its parent $\Lambda$ ($\overline{\Lambda}$). In the sense, the hyperon's decay is its own polarimeter. It is worth mentioning that this prescription is not valid if in QM since it ignores the contribution from interference between the two spin states. In QM, the entire process $\mathrm{c\overline{c} \rightarrow \Lambda \overline{\Lambda} \rightarrow p \pi p \pi}$ should be treated as a coherent process in which the spins of the intermediate state (hyperons in this case) cannot in principle be well-defined. \\
While Tornqvist suggested to test QM with the use of this tool \cite{Ton1981} \cite{Ton1986}, S. P. Baranov extended it for testing local realistic theory (LRT) by reformulating BI with respective to the orientation of decay product p and $\mathrm{\overline{p}}$ in the final state \cite{Baranov} with the help of (\ref{eq1}). Different formulation in similar channels ($\eta \rightarrow VV$) is discussed by J. Li {\it et. al.} \cite{etaVV}. In this paper we develop further formulation and give a comprehensive discussion. Specifically: \\

\begin{itemize}
\item Transform BI into a representation in terms of the direction of decayed particles.
\item The evaluation of whether $\mathrm{c\overline{c} \rightarrow \Lambda \overline{\Lambda} \rightarrow p \pi p \pi}$ has sensitivity of testing this newly derived BI by QM-based calculations.
\item An analysis of achievable significance of such a test and its experimental feasibility. 
\end{itemize}

\phantom{MM} 
\phantom{MM} 
\phantom{MM}

\begin{figure}[H]
\begin{center}
\includegraphics[width=14.5cm]{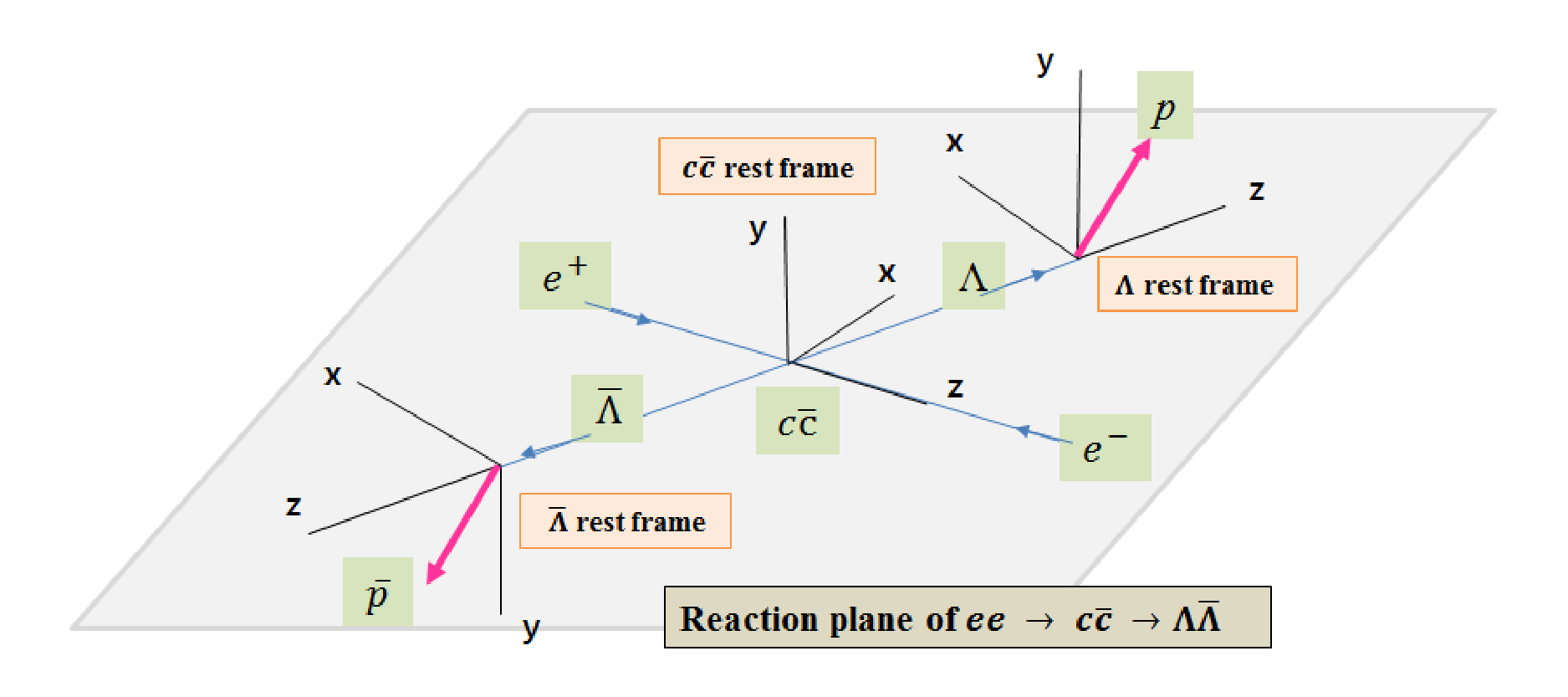}
\caption{An overview of the $c \overline{c} \rightarrow \Lambda \overline{\Lambda} \rightarrow \mathrm {p \pi^- \overline{p} \pi^+}$ process. First, a charm meson decays into a $\Lambda \overline{\Lambda}$ pair. These travel back-to-back in the meson's rest frame and decay into p$\pi^-$ and $\mathrm{\overline{\mathrm{p}}}$$\pi^+$ respectively. We measure the orientation of p and $\overline{\mathrm{p}}$ as the testing variables of BI. Experimentally, the charm mesons are supposed to be produced in $e^+e^-$ collisions. J/$\psi$ is obtained directly while $\eta_c$ and $\chi_{c0}$ are generated via the decay of J/$\psi$ and $\psi'$ respectively.}
\label{fig.1}
\end{center}
\end{figure}
\begin{figure}[H]
\begin{center}
\label{fig.2}
\includegraphics[width=8.3cm]{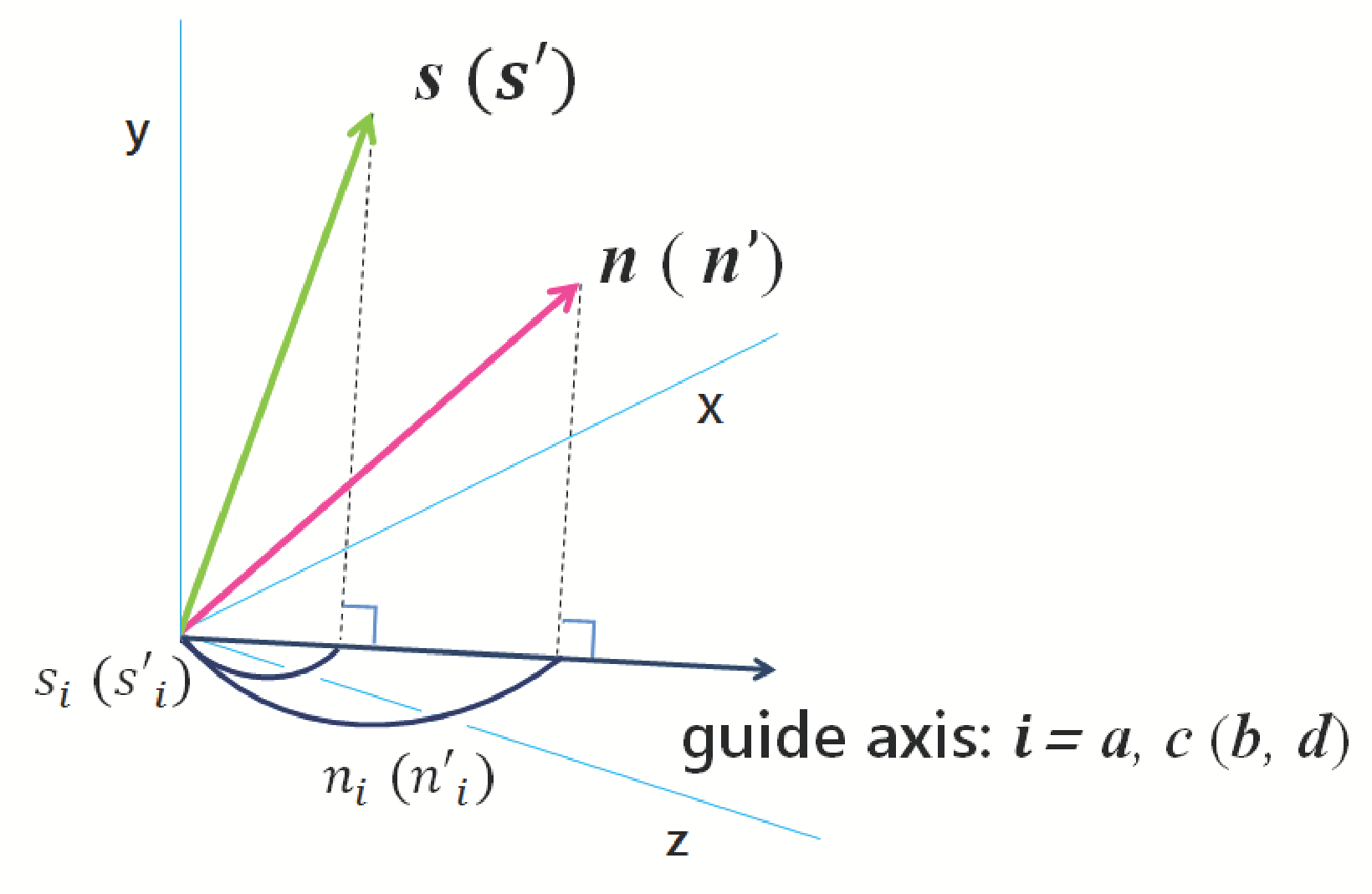}
\caption{The relation of unit vectors defined in the $\Lambda$ ($\bar{\Lambda}$) rest frame and their projections. $\vec{n}$, ($\vec{n'}$) is the orientation of outgoing proton (anti-proton), $\vec{s}$ ($\vec{s'}$) the polarization of $\Lambda$ ($\bar{\Lambda}$). Their projection onto a guide axis is labeled by the attached index.}
\end{center}
\end{figure}
%
%
%
%
\section{Momentum representation of Bell's inequality}
In realistic theories, a particle's spin is a definite physical quantity and normally treated as a 3 dimensional continuously valued vector just like classical angular momentum. Consider a two particles system with their spins of one half. The polarization vector $\bm{s}$ and $\bm{s}'$ follow the algebraic condition: 
\begin{equation}
| \braket{ \, s_{a} \, s'_{b} \, } - \braket{ \, s_{a} \, s'_{c} \, }  |  \,  \leq \, 1+ \braket{ \, s_{b} \, s'_{c} \, }.
\label{eq.bell}
\end{equation}
where $\vec{a}$, $\vec{b}$, and $\vec{c}$ are arbitrary unit vectors ``guide axes''. $\bm{s}$ and $\bm{s}'$ have the norms of 1 and $s_{a}$ is its projection onto $\vec{a}$ i.e. $s_a = \bm{s} \cdot \bm{a}$; $s'_{b}$ is that of particle 2 onto $\vec{b}$; $s_{b}$, $s'_{c}$ defined similarly (Fig. 2). Here $\bm{s}$ and $\bm{s}'$ are quoted as the realistic value of polarizations and considering the case in that they distribute probabilistically, and the ensemble average is weighted by the probability density function of them. These are seemingly the striking deferences from the conventional BI \cite{Bell} which has the exactly same form as (\ref{eq.bell}),
\begin{equation}
| \braket{ \, m_{a} \, m'_{b} \, } - \braket{ \, m_{a} \, m'_{c} \, }  |  \,  \leq \, 1+ \braket{ \, m_{b} \, m'_{c} \, }
\label{eq.cbell}
\end{equation}
with $m_i$ and $m'_i$ being the result of measuring $s_i$ and $s'_i$ (i=a, b, c). (\ref{eq.cbell}) is valid no matter if $m_i$ ($m'_i$) is continuous or discrete providing $-1 \leq m_i \, (m'_i) \leq 1$, or if the realistic values $s_i$ ($s'_i$) are different by each measurement, though the original Bell's discussion \cite{Bell} takes discrete ones assuming measuring the same spin by a Stern-Gerlach type of experiment. Thus, since the both inequalities (\ref{eq.bell}) and (\ref{eq.cbell}) has the same essence that they generally hold in (local) realistic view of physics and the constraints are due to the realistic interpretation of spin, we call the inequality (\ref{eq.bell}) BI as well and use it to test LRTs. \\
%
%
In extending to a relativistic case, in which the particles 1 and 2 belong to different respective frames, the CHSH (Clauser-Horne-Shimony-Holt) version of BI gives a more appropreate description \cite{CHSH}:
\begin{equation}
| \braket{ \, s_{a} \, s'_{b} \, \, } + \braket{ \, s_{a} \, s'_{d} \, } + \braket{ \, s_{c} \, s'_{b} \,} - \braket{ \, s_{c} \, s'_{d} \,}  |  \,  \leq \, 2
\label{eq.chsh}
\end{equation}
Here four guide axes are involved, two of which ($\vec{a}$, $\vec{c}$) are used for particle 1 and the other two ($\vec{b}$, $\vec{d}$) for particle 2. Each set of two guide axes is defined independently in their own frame. Recall that the original version of BI (\ref{eq.bell}) has $\vec{b}$ in common to both frames. This leads to a confusion when in a relativistic case, as pointed out by \cite{Baranov}, while the CHSH inequality (\ref{eq.chsh}) remains well-defined with no such ambiguity. \\
At this point, it is clear that (\ref{eq.chsh}) is not capable of being tested by experiment because it is denoted by the realistic values instead measured values, that can yield different values due to the disturbance by the involvement of hidden variables, generally in LRTs. However we can translate it into a testable one, assuming the angular distribution (\ref{eq1}) for the decay $\mathrm{\Lambda \rightarrow p\pi^- }$ ($\mathrm{\overline{\Lambda} \rightarrow \overline{\mathrm{p}}\pi^+}$). If the two decays are independent, the correlations of hyperon spin and the proton (anti-proton) orientation can be tagged as
\begin{equation}
\braket{ \, s_{a} \, s'_{b} \,} = - \frac{9}{\alpha_{\Lambda}\alpha_{\bar{\Lambda}}} \braket{ \, (\vec{n} \cdot \vec{a}) (\vec{n'} \cdot \vec{b}) \,} =: - \frac{9}{\alpha_{\Lambda}\alpha_{\bar{\Lambda}}} \braket{ \, n_a \, n'_b \,} .
\label{eq.4}
\end{equation}
$\bm{n}$ ($\bm{n}'$) is the traveling directions of the proton (anti-proton). The equation was originally provided by S. P. Baranov \cite{Baranov}, which however takes a different picture of classical spin reality from one in this paper. The derivation for our spin interpretation is given in the appendix, yielding the same result as \cite{Baranov}. Equation (\ref{eq.chsh}) can therefore be written as 
\begin{table}[t]
	\begin{center}
		\caption{Kinematics of $\Lambda$ and $\overline{\Lambda}$ from each charmonium decay. $\omega$ is the fraction that the two decays are in space-like configuration: $\omega = 2 \,  \int_{0}^{\infty} dt_1 \int_{t_1}^{\frac{1+\beta}{1-\beta} t_1} dt_2 \, \, \frac{1}{\tau} e^{-\frac{t_1}{\tau}} \frac{1}{\tau} \, e^{-\frac{t_2}{\tau}} = \beta $.}
			\begin{tabular}{ c | c c c   }
            		\multicolumn{4}{c}{} \\
			\hline
	$$ & $\eta_c$ & $\chi_{c0}$ & $J/\psi$  \\
	\hline
	$\beta$ & $0.663$ & $0.757$ & $0.693$ \\
	$\gamma$ & $1.329$ & $1.621$ & $1.408$ \\
	$\omega$ & $0.663$ & $0.757$ & $0.693$ \\
		\hline
		\end{tabular}
                  \label{tab.1}
	\end{center}
\end{table}
\begin{equation}
|\braket{ \,n_a \, n'_b \, } + \braket{ \,n_a \, n'_d \,} + \braket{ \,n_c \, n'_b \,} - \braket{ \,n_c \, n'_d \,}  |  \,  \leq \,  \frac{2\alpha_{\Lambda}\alpha_{\bar{\Lambda}}}{9}
\label{eq.5}
\end{equation} 
We call this ``the momentum representation'' of BI. Here $\bm{n}$ and $\bm{n}'$ are measured values, in addition, have corresponding observables in QM. Therefore (\ref{eq.5}) is eventually an inequality which can be evaluated by experiment and by QM.   \\ \\

Several comments should be added to this reformulation: 
\begin{itemize}
%
%
%
\item On deriving (\ref{eq.4}),  we assumed that the decay of the first hyperon only depends on the polarization of that particular hyperon. It seems reasonable according to the kinematical properties of  $\Lambda$ and $\overline{\Lambda}$ from each meson decay, as shown in Table 1. The two decays have a space-like separation for 66 $\% \sim$ 76 $\%$ of events, across which no interaction can act. Experimentally, these space-like events can be selectively extracted, which realizes a complete isolation of the decays. \\
\item In a conventional tests with direct spin measurement, guide axes play much important physical role in that only one direction of spin can be chose to be measured. In our test, however, guide axes are just arbitrary unit vectors as seen in (\ref{eq.bell}). This cause no problem because momentum components can be simultaneously determined within one measurement, and even has an advantage in that we will not suffer from the ``free will'' problem in choosing guide axes. \\

\item On the other hand, it has no scheme comparable to ''delayed-choice measurement'' in optical experiment which prevents the hyperons from deciding the way of decaying in advance, or exchanging information each other just after their production so that they can have a larger correlation in their orientations. These are expected to be the loopholes of this test. \\

\item The development from (\ref{eq.chsh}) to (\ref{eq.5}) proceeds based wholly on the classical picture. Although (\ref{eq1}) can be derived by a standard QM calculation, we accept (\ref{eq1}) as just an experimental fact without assuming any background theories underlying it. \\

\item There has been a claim that BI with respect to commuting variables can not violate even in QM  due to their incoherency \cite{Abel}. However this is not the case because the inequality (\ref{eq.5}) is essentially a BI with respect to spins. We just transform and represent it with $\bm{n}$ and $\bm{n'}$, as a result, the upper limit of the BI is lower by factor of $\frac{\alpha_{\Lambda} \alpha_{\bar{\Lambda}} }{9}$ ($\sim \frac{1}{20}$) than that of a BI which naively takes variables as $\bm{n}$ ($\bm{n'}$): $|\braket{ \,n_a \, n'_b \, } + \braket{ \,n_a \, n'_d \,} + \braket{ \,n_c \, n'_b \,} - \braket{ \,n_c \, n'_d \,}  |  \,  \leq \,  2$. Thus the BI (\ref{eq.5}) does violate, as we show in the section 4. \\
%

\item It may seem to be against the intuition that (\ref{eq.5}) is testable inequality regardless of the magnitude of $\alpha_{\Lambda}$ which represents the extent of the correlation between $\bm{s}$ and $\bm{n}$. Theoretically this is true since the upper limit (the right-hand side of (\ref{eq.5}) drops accordingly. The conceptional essence of the test is not how much $\bm{n}$ ($\bm{n'}$) acts as the polarimeter of hyperons but that the correlations $\braket{ \,s_i \, s'_j \,}$ and $\braket{ \,n_i \, n'_j \,}$ have an exact relation (\ref{eq.4}) thus we can set the upper limit on $\braket{ \,n_i \, n'_j \,}$ as (\ref{eq.5}). Practically, however, the small correlation amplitude of $\braket{ \,n_i \, n'_j \,}$ is difficult to measure precisely, therefore we at last prefer the decay channel with as large $\alpha_{\Lambda}$ as possible. \\

\end{itemize} 
%
%
%
\section{Bilinear expression of the momentum represented BI (\ref{eq.5})}
For later analysis, it is convenient to transform the BI (\ref{eq.5}) into one written with a ``correlation matrix'' $\hat{C}$. $\hat{C}$ is a 3$\times$3 real valued matrix given by the correlation amplitude, 
\begin{eqnarray*}
\hat{C}_{ij}  := \braket{ \, n_i n'_j \,} \mbox{\phantom{MMMM}} (i,\, j = 1,2,3)
\end{eqnarray*}
where the indices $i$, $j$ label $x$, $y$, $z$ components of a vector in Cartesian coordinate. 
$\braket{ \, n_a \, n'_b \, }$ can be written in a simple form of bilinear using matrix $\hat{C}$ as
\begin{equation}
\braket{ \,  n_a \, n'_b  \, } \,\, =  \,\, \sum_{i=1}^3 \sum_{j=1}^3  a_i \, b_j \braket{ \, n_i \, n
'_j \,}  \,\, = \,\, \vec{a}^T \hat{C} \, \vec{b}   \,\, .                  
\label{eq.6}
\end{equation}
We now define $\hat{Q}$ as the left-hand side of (\ref{eq.5}), which can be written as a sum of bilinears:
\begin{equation}
{ \hat{Q} } =  | \,  \vec{a}^T \hat{C} \, {(\vec{b}+\vec{d})} +  \vec{c}^T \hat{C} \, {(\vec{b}-\vec{d})} \, | .
\label{eq.7}
\end{equation}
The advantage of this representation can be seen in that the physical part ($\hat{C}$) and physics-independent part (guide axes) are well separated. \\
Next we try to specify the maximum value of $\hat{Q}$ and to get rid of guide axes out of the in equality since they give no physical importance in our formulation. It is easy to show that the maximum value of $\hat{Q}$ is same as that of $\vec{a}^T \hat{C} (\vec{b}+\vec{d})+ {\vec{c}^T \hat{C}(\vec{b}-\vec{d}})$, thus considering the case of $\hat{Q} = \vec{a}^T \hat{C} (\vec{b}+\vec{d})+ {\vec{c}^T \hat{C}(\vec{b}-\vec{d})}$ is sufficient. The method of Lagrange multipliers (MLM) can be utilized. With the constraint conditions $\vec{a}^T \vec{a} = 1, \vec{b}^T \vec{b} = 1, \vec{c}^T \vec{c} = 1$ and $\vec{d}^T \vec{d} = 1$, a scalar function $L$ can be constructed using four multipliers $\xi_{a}$, $\xi_{b}$, $\xi_{c}$ and $\xi_{d}$ .
\begin{eqnarray*}
L & = & \vec{a}^T \hat{C} \, (\vec{b}+\vec{d})+ \vec{c}^T \hat{C} \, (\vec{b}-\vec{d}) \\
 & - & \frac{1}{2} \xi_{a} \, (\vec{a}^T \vec{a} - 1) - \frac{1}{2} \xi_{b} \, (\vec{b}^T \vec{b} - 1) - \frac{1}{2} \xi_{c} \, (\vec{c}^T \vec{c} - 1) - \frac{1}{2} \xi_{d} \, (\vec{d}^T \vec{d} - 1)
\end{eqnarray*}
Setting all the derivatives to zero,
$$
\mbox{\phantom{.}} \frac{\partial L}{\partial \vec{a}^T}  =  0 \mbox{\phantom{MM}} \iff \mbox{\phantom{MkM..}} { \hat{C}(\vec{b} + \vec{d}) \, - \, } \xi_a  {\vec{a} = 0 } \label{eq.8a} \eqno{(8a)} 
$$
$$
\mbox{\phantom{k}} \frac{\partial L}{\partial  \vec{b}} \, = 0 \mbox{\phantom{MM}}  \iff \mbox{\phantom{Mlll}} (\vec{a} + \vec{c})^T \hat{C} -  \xi_b \vec{b}^T = 0  \label{eq.8b} \eqno{(8b)}
$$
$$
\mbox{\phantom{.}} \frac{\partial L}{\partial \vec{c}^T} = 0  \mbox{\phantom{MM}}  \iff \mbox{\phantom{MMM..}} { \hat{C}(\vec{b} - \vec{d}) \, - \, } \xi_c  {\vec{c} = 0 } \label{eq.8c} \eqno{(8c)} 
$$
$$
\mbox{\phantom{..}} \frac{\partial L}{\partial \vec{d}} \,  = 0 \mbox{\phantom{MM}}  \iff \mbox{\phantom{Mk.}} (\vec{a} - \vec{c})^T \hat{C} -  \xi_d \vec{d}^T = 0  \label{eq.8d} \eqno{(8d)} 
$$
$$
\mbox{\phantom{MMkkMMMMMM}} \frac{\partial L}{\partial \xi_{\alpha}} = 0  \mbox{\phantom{MM}} \iff \mbox{\phantom{M}(constraint conditions)} \mbox{\phantom{MMM}} (\alpha = a, b, c, d) .
\label{eq.9} \eqno{(9)}
$$
\setcounter{equation}{9}
Mutiplying $\vec{a}^T,  \vec{c}^T (  \vec{b}, \vec{d})$  from the left (right)-hand side of (8a) $\sim$ (8d) respectively and using the constraint conditions (9), the multipliers $\xi_{\alpha}$ can be written as
\begin{eqnarray}
 \xi_a =   \vec{a}^T \hat{C}(\vec{b}+\vec{d})  \mbox{\phantom{MMMMMMM}}
 \xi_c =   \vec{c}^T \hat{C}(\vec{b}-\vec{d})  \nonumber \\
 \xi_b =   (\vec{a}+\vec{c})^T \hat{C}\vec{b}  \mbox{\phantom{MMMMMMM}}
 \xi_d =   (\vec{a} - \vec{c})^T \hat{C}\vec{d}  
 \label{eq.10}
\end{eqnarray}
Substituting these back to (8b) (8d) gives
\begin{eqnarray}
\vec{b} = \frac{{ \hat{C}^T (\vec{a}+\vec{c}) } }
{(\vec{a}+\vec{c})^T \hat{C}\vec{b} }
=: \frac{  \hat{C}^T (\vec{a}+\vec{c})   } {\rho} \nonumber \\
\vec{d} = \frac{{ \hat{C}^T (\vec{a}-\vec{c}) } }
{  {(\vec{a}-\vec{c})^T \hat{C}\vec{d} }}
=:  \frac{ {\hat{C}^T (\vec{a}-\vec{c}) }  } {\sigma} .
\label{eq.11}
\end{eqnarray} \\
Note that $\vec{b}$ and $\vec{d}$ are parallel to $ {\hat{C}^T (\vec{a} + \vec{c})}$ and $ {\hat{C}^T (\vec{a} - \vec{c})}$. Normalization factors $\rho$ and $\sigma$ are chosen to satisfy $ {\vec{b}^T \vec{b}} = 1$ and $ {\vec{d}^T \vec{d}} = 1$ i.e.
\begin{eqnarray}
\rho = \sqrt{ { (\vec{a}+\vec{c}) \hat{C}^T  \hat{C} (\vec{a}+\vec{c}) }} \mbox{\phantom{MMMMM}} 
\sigma = \sqrt{ { (\vec{a}-\vec{c}) \hat{C}^T  \hat{C} (\vec{a}-\vec{c}) }}
\label{eq.12}
\end{eqnarray}
%
%
Now we define a symmetric matrix $S$ as ${S := \hat{C}^T  \hat{C} = \hat{C} \hat{C}^T}$. With (\ref{eq.10}) (\ref{eq.11}), we eliminate $\vec{b}$ and $\vec{d}$ in (8a)(8c), giving
\begin{eqnarray}
(\sigma + \rho) S \vec{a} + (\sigma - \rho) S \vec{c} & = & \mu \vec{a}  \nonumber \\
(\sigma - \rho) S \vec{a} + (\sigma + \rho) S \vec{c} & = & \nu \vec{c}  \nonumber 
\end{eqnarray}
where
\begin{eqnarray}
\mu & := & { ( \sigma + \rho) \vec{a}^T S \vec{a} + (\sigma - \rho) \vec{a}^T S \vec{c} } \nonumber \\
\nu  & := & { ( \sigma + \rho) \vec{c}^T S \vec{a} + (\sigma - \rho) \vec{c}^T S \vec{c} }\,\, . \nonumber
\end{eqnarray}
These can be expressed as eigen-equations for the vectors $\vec{a}$ and $\vec{c}$
\begin{eqnarray}
{  \left[    -4\rho \sigma S^2 + ( \sigma + \rho) (\mu+\nu) S \right] \vec{a} } & = & { \mu \nu} \vec{a} \nonumber \\
{  \left[     4 \rho \sigma S^2 + ( \sigma + \rho) (\mu+\nu) S \right] \vec{c} } & = & { \mu \nu} \vec{c} .\label{eq.13}
\end{eqnarray}

It can be easily verified that the matrices appearing in the left-hand sides of (\ref{eq.13}) $\pm 4\rho \sigma S^2 + ( \sigma + \rho) (\mu+\nu) S$ have identical eigen-vectors to those of $S$ as 
\begin{eqnarray*}
\left[\,  \pm 4\rho \sigma {S}^2 + ( \sigma + \rho) (\mu+\nu) {S} \, \right] \vec{v}_i = \left[ \, \pm 4\rho \sigma \lambda_i^2 + ( \sigma + \rho) (\mu+\nu) \lambda_i \, \right] \vec{v}_i  
\end{eqnarray*}
with $\lambda_i$, $\vec{v}_i$ being the eigen-values and eigen-vectors satisfying ${ S \vec{v}}_i = \lambda_i {\vec{v}}_i$.
Since any 3-dimensional matrix can only have three independent eigen-vectors at most, $\vec{v}_i (i=1,2,3)$ give a full description of all solutions of (\ref{eq.13}). Therefore,
\begin{eqnarray}
&& \vec{a} = \pm \vec{v}_i \nonumber \\
&& \vec{c} = \pm \vec{v}_j  \nonumber \\
&& |\vec{v}_i| = 1 \mbox{\phantom{MMM} } (i,\, j = 1,2,3)
\label{eq.14}
\end{eqnarray}
are required and we see all of these satisfy (\ref{eq.13}). Note that the norms of $\vec{a}$ and $\vec{c}$ are confirmed to be 1. Substituting these into (\ref{eq.11})(\ref{eq.12}), $\vec{b}$, $\vec{d}$ and all other coefficients are determined. Setting $( \vec{a}, \, \vec{c}) = (\vec{v_i}, \,  \vec{v_j})$ for simplicity,
\begin{eqnarray}
\rho  = \sqrt{(\lambda_i + \lambda_j)  ( 1+ \vec{v}_i^T \cdot \vec{v}_j )} \nonumber \\
\sigma = \sqrt{(\lambda_i + \lambda_j)  (1- \vec{v}_i^T \cdot \vec{v}_j )} \nonumber 
\end{eqnarray}

$ {(\mathrm{i}) \,\, \vec{a}} = \vec{c} \,\, (i = j) $
\begin{eqnarray}
&& \rho  =  2 \sqrt{\lambda_i} \nonumber \mbox{\phantom{MMMM}}
\sigma  =  0 \nonumber \\
&& \vec{b}  = \frac{  \hat{C}^T \vec{v}_i }{ \sqrt{ \lambda_i}} \mbox{\phantom{MMMM}}
\vec{d}  = (\mbox{arbitrary unit vector}) \nonumber \\
&& {\hat{Q}} = 2\vec{a}^T \hat{C} \vec{b} \,\, = \,\, 2\sqrt{\lambda_i}   
\label{eq.15-1}
\end{eqnarray}

${(\mathrm{ii}) \,\, \vec{a}} \neq \vec{c} \,\, (i \neq j)$
\begin{eqnarray}
&& \rho = \sigma = \sqrt{\lambda_i + \lambda_j} \nonumber \\
&& \vec{b} = \frac{  \hat{C}^T (\vec{v}_i + \vec{v}_j)}{ \sqrt{\lambda_i + \lambda_j} } \nonumber \mbox{\phantom{MMMM}}
 \vec{d} = \frac{  \hat{C}^T (\vec{v}_i - \vec{v}_j)}{ \sqrt{\lambda_i + \lambda_j} } \nonumber \\
&& {\hat{Q}} = 2\sqrt{\lambda_i + \lambda_j} \label{eq.15}
\end{eqnarray}
In deriving (\ref{eq.15}) we used the fact that ${S = \hat{C}^T \hat{C}}$ is a symmetric matrix with orthogonal  eigen-vectors (${\vec{v}}_i^T \cdot {\vec{v}}_j = \delta_{ij}$). The same analysis can be applied for the case of $( \vec{a}, \, \vec{c}) = (\vec{v_i}, \, - \vec{v_j}), \, (- \vec{v_i}, \, \vec{v_j}), \, (- \vec{v_i}, \, - \vec{v_j}) $. The maximum $\hat{Q}$ value comes from (\ref{eq.15}) when we set ${\lambda_i}$ and ${\lambda_j}$ to the largest two eigen-values of $S$. One point to be noted is that solutions of MLM generally include local maxima where the real maximum value is given not at extrema but at the boundary of parameter space. However this is not the case here, since the parameter space of $\vec{a}$, $\vec{b}$, $\vec{c}$, and $\vec{d}$ are independent ``spheres'' that have no boundary.\\

With all these considerations, we finally reach the form,

\begin{equation}
\hat{Q}_{\mathrm{max}} = 2\sqrt { \lambda_1 + \lambda_2 }  \leq  \frac{2\alpha_{\Lambda}\alpha_{\bar{\Lambda}}}{9}
\label{eq.16}
\end{equation}
where ${\lambda_1}$ and ${\lambda_2}$ are the largest two eigen-values of ${\hat{C}^T \hat{C}}$. For later convenience, we define $C$ and $Q$ by dividing by the right-hand side of (\ref{eq.16})

\begin{eqnarray}
&& C_{ij} = \braket{ \, n_{i} \, n'_{j} \,} \frac{9}{2\alpha_{\Lambda}\alpha_{\bar{\Lambda}}}    \mbox{\phantom{MMMM}}      (i,j=1,2,3)  \nonumber \\      
&&  Q_{\mathrm{max}} =  2 \sqrt { \lambda_1 + \lambda_2 } \mbox{\phantom{MMMMM}}       \mbox{($\lambda_1$, $\lambda_2$: the largest two eigen-values of ${ C^T C }$) } \nonumber \\
&& Q_{\mathrm{max}} \leq 1.                                                     
\label{eq.16-2}
\end{eqnarray}
This (\ref{eq.16-2}) is our target expression. The classical limit of $Q_{\mathrm{max}}$ is ${Q_{\mathrm{CL}}} = 1$ while the quantum limit reaches $\sqrt{2}$. \\ \\

\section{Quantum mechanical calculation of $C$ and $Q_{\mathrm{max}}$}
In this section, we perform a QM-based computation of the correlation matrix $C$ and $Q_{\mathrm{max}}$ for each channel $\mathrm{\eta_c \rightarrow \Lambda\overline{\Lambda} \rightarrow p\pi^-\overline{\mathrm{p}}\pi^+}$, $\mathrm{\chi_{c0}
 \rightarrow \Lambda\overline{\Lambda} \rightarrow p\pi^-\overline{\mathrm{p}}\pi^+}$ and $\mathrm{J/\psi \rightarrow \Lambda\overline{\Lambda} \rightarrow p\pi^-\overline{\mathrm{p}}\pi^+}$, to determine if the BI (\ref{eq.16-2}) is held or violated in QM. \\
The matrix elements for each channel are as below, according Feynman's rules and the effective  Lagrangian prescription. 
\begin{eqnarray}
\mathcal{M}_{\eta_c} \,\,\, & = & \mathcal{M}_{\Lambda} (\vec{p}_{\Lambda}, s_{\Lambda}, \vec{p}_{\mathrm{p}}, s_{\mathrm{p}}) \, \bar{u}(\vec{p}_{\Lambda}, s_{\Lambda}) \, \gamma^{5} \, v(\vec{p}_{\bar{\Lambda}}, s_{\bar{\Lambda}}) \, \mathcal{M}_{\bar{\Lambda}}( \vec{p}_{\bar{\Lambda}}, s_{\bar{\Lambda}},  \vec{p}_{\bar{\mathrm{p}}}, s_{\bar{\mathrm{p}}}) \nonumber \\
\mathcal{M}_{\chi_{c0}} \, & = & \mathcal{M}_{\Lambda} (\vec{p}_{\Lambda}, s_{\Lambda}, \vec{p}_{\mathrm{p}}, s_{\mathrm{p}}) \, \bar{u}(\vec{p}_{\Lambda}, s_{\Lambda})  \, v(\vec{p}_{\bar{\Lambda}}, s_{\bar{\Lambda}}) \, \mathcal{M}_{\bar{\Lambda}}( \vec{p}_{\bar{\Lambda}}, s_{\bar{\Lambda}},  \vec{p}_{\bar{\mathrm{p}}}, s_{\bar{\mathrm{p}}}) \nonumber \\
\mathcal{M}_{\mathrm{J/_\psi}} & \propto & \mathcal{M}_{\Lambda} (\vec{p}_{\Lambda}, s_{\Lambda}, \vec{p}_{\mathrm{p}}, s_{\mathrm{p}}) \, \bar{u}(\vec{p}_{\Lambda}, s_{\Lambda}) \, \epsilon_{\mu} \left[ \gamma^{\mu} + \frac {a_{\psi}}{m_{\psi}} (p_{\Lambda}^{\mu} - p_{\bar{\Lambda}}^{\mu}) \right] \, v(\vec{p}_{\bar{\Lambda}}, s_{\bar{\Lambda}}) \, \mathcal{M}_{\bar{\Lambda}}( \vec{p}_{\bar{\Lambda}}, s_{\bar{\Lambda}},  \vec{p}_{\bar{\mathrm{p}}}, s_{\bar{\mathrm{p}}}) \nonumber \\ \nonumber \\
&& \mathcal{M}_{\Lambda} (\vec{p}_{\Lambda}, s_{\Lambda}, \vec{p}_{\mathrm{p}}, s_{\mathrm{p}}) = \bar{u}(\vec{p}_p, s_{\mathrm{p}}) \, ( 1+ c_{\Lambda} \gamma^5) \,u(\vec{p}_{\Lambda}, s_{\Lambda}) \nonumber \\
&& \mathcal{M}_{\bar{\Lambda}}( \vec{p}_{\bar{\Lambda}}, s_{\bar{\Lambda}},  \vec{p}_{\bar{\mathrm{p}}}, s_{\bar{\mathrm{p}}}) = \bar{v}(\vec{p}_{\bar{\Lambda}}, s_{\bar{\Lambda}}) \, ( 1- c_{\bar{\Lambda}} \gamma^5) \, v(\vec{p}_{\bar{\mathrm{p}}}, s_{\bar{\mathrm{p}}})  \nonumber
\label{eq.17}
\end{eqnarray}
$s_A$ denotes the helicity of particle A, $u(\vec{p}_A, s_A)$ and $v(\vec{p}_A, s_A)$ the 4-spinor and $\epsilon^{\mu}$ the polarization vector of the meson J/$\mathrm{\Psi}$.  Momenta appearing here $\vec{p}_A$, $p^{\mu}_A$ are all defined in the \underline{decaying meson rest frame}. $\mathcal{M}_{\Lambda}$ and $\mathcal{M}_{\bar{\Lambda}}$ are the matrix elements responsible for the $ \Lambda \rightarrow \mathrm{p} \pi^-$ and $\overline{\Lambda} \rightarrow \mathrm{\overline{p}} \pi^+ $ sector which gives $c_{\Lambda}$ = $c_{\bar{\Lambda}}$ providing CP conservation argument. The distribution of $\Lambda$ decay and the decay parameter $\alpha_{\Lambda}$ in (\ref{eq1}) is associated as
\begin{align}
\frac{d \, \Gamma_{\Lambda}}{\, d\Omega_{\mathrm{p}} }  & \propto \sum_{s_{\mathrm{p}}} \left | \mathcal{M}_{\Lambda} \right |^2 \\
\alpha_{\Lambda} & = \frac{-2| \vec{p}| c_{\Lambda}}{E(1+c_{\Lambda}^2) + m_{\mathrm{p}}(1-c_{\Lambda}^2)},
\end{align}
with which the parameter $c_{\Lambda}$ is derived to $c_{\Lambda} = c_{\bar{\Lambda}} = -6.79 \pm 0.18$. $E$, $\vec{p}$ and $m_p$ are the energy, momentum of outgoing proton in the $\Lambda$ rest frame and its rest mass respectively. \\
$\mathcal{M}_{\mathrm{J/_\psi}}$ includes an additional parameters $a_{\psi}$ associated with the form factor of J/$\mathrm{\psi}$. This has been experimentally determined from the unpolarized $\Lambda$ distribution of $e^+e^- \rightarrow J/\psi \rightarrow \Lambda \overline{\Lambda}$ where $e^+e^-$ act as chiral fermions leading $J/\Psi$ polarization parallel or anti-parallel to the beam axis according to the helicity conservation in a high energy system:
\begin{align}
& \frac{d \, \Gamma_{J/\Psi}}{\, d\Omega_\mathrm{\Lambda} } 
	\propto \sum_{s_{\mathrm{\Lambda}}, s_{J/\Psi}} \left | \mathcal{M}_{J/\Psi} \right |^2 
	=  1 + F \cos^2\theta  \nonumber \\
& F =  \frac{(M^2 - 4m^2)(1-r^2)}{ (1+r^2)(M^2 + 4m^2) - 8Mmr }  \mbox{\phantom{MMM}} r :=  \frac{a_{\psi}}{\frac{2m}{M} a_{\psi} + 1}
\end{align}
$M$ and $m$ are the masses of J$/\psi$ and $\Lambda$ respectively. $\theta$ is the angle between the $e^+e^-$ beam axis and the direction of $\Lambda$. The observed value of $F$ is $F = 0.65 \pm 0.11$ \cite{BES2} which gives two possible solutions $a_{\psi} = 2.46 \pm 0.18$ and $ a_{\psi}= 0.54 \pm 0.18$. In our calculation here, both lead to identical results. (It probably depends only on $F$.) \\

The angular distribution for the coherent process $\mathrm{c\bar{c}} \rightarrow \mathrm{\Lambda\overline{\Lambda}} \rightarrow {p \pi p \pi}$ is therefore obtained, assuming unpolarized initial and final states,
\begin{equation}
\left( \frac{d\sigma}{d\Omega_{\Lambda}d\Omega_\mathrm{p} d\Omega_{\bar{\mathrm{p}}}} \right)  \,\, \propto \mathrm{ave} \left| \sum_{s_{\Lambda}, s_{\bar{\Lambda}}}\mathcal{M} \, \right| ^2
=: \overline{\, \left| \mathcal{M}_0 \right|^2 }  .
\end{equation}
$d\Omega_\mathrm{p}$ ($d\Omega_{\bar{\mathrm{p}}}$) is the solid angle element in the proton (anti-proton) momentum space in  $\Lambda$ ($\overline{\Lambda}$) rest frame. The average is taken over the spins of particles in initial and final states. Note that the sums over $s_{\Lambda}$ and $s_{\bar{\Lambda}}$ are taken before squaring, which leads to the terms representing quantum interference between the two intermediating spin states. The correlation matrix $C$ is calculated as 
%
\begin{eqnarray}
C_{ij} & = & \braket{ \, n_i \, n'_j \, }
 \frac{9}{2\alpha_{\Lambda}^2} \nonumber \\ 
& = & \frac{\braket{ \, p_i \, p'_j \, } }{|\vec{p}|^2}
 \frac{9}{2\alpha_{\Lambda}^2} \nonumber \\ 
& = & \frac {9} {2\alpha_{\Lambda}^2 |\vec{p}|^2}   
\int d\Omega_{\Lambda} d\Omega_\mathrm{p} \, d\Omega_{\bar{\mathrm{p}}}
 \, \, p_i \, p'_j \, \, \mathrm{Prob(\vec{p}, \vec{p} ', \vec{p}_\Lambda}) \nonumber \\
& = & \frac {9} {2\alpha_{\Lambda}^2 |\vec{p}|^2}  
\int d\Omega_{\Lambda} d\Omega_\mathrm{p} \, d\Omega_{\bar{\mathrm{p}}}
\, \, p_i \, p'_j \, \, \left( \frac{d\sigma}{d\Omega_{\Lambda} d\Omega \, d\Omega_{\bar{\mathrm{p}}}}/\sigma_{tot} \right)  \nonumber \\
& = & \frac {9} {2\alpha_{\Lambda}^2 |\vec{p}|^2} \frac{\int d\Omega_{\Lambda} d\Omega_\mathrm{p} \, d\Omega_{\bar{\mathrm{p}}} \, \, p_i \, p'_j \, \overline{\, \left| \mathcal{M}_0 \right|^2 }}  {\int d\Omega_{\Lambda} d\Omega_\mathrm{p} \, d\Omega_{\bar{\mathrm{p}}} \, \, \overline{\, \left
|\mathcal{M}_0 \right|^2 }}  \\
\nonumber
\label{eq.18}
\end{eqnarray}
%
%
$\vec{p}$ ($\vec{p'}$) is defined as the proton (anti-proton) momentum in the $\Lambda$ ($\overline{\Lambda}$) rest frame. (Recall that momenta $\vec{p_p}$ ($\vec{p_{\bar{p}}})$ used in computing $\overline{\, \left| \mathcal{M}_0 \right|^2 }$ above is one in the decaying meson rest frame.) 
The result of $C_{ij}$ and $Q_{\mathrm{max}}$ for each channel is shown in Table 2. The off-diagonal components are always 0, reflecting the symmetry in the processes, thus $Q_{\mathrm{max}} =  2 \sqrt{C_{11}^2+C_{33}^2}$ following (\ref{eq.16-2}). \\ 
\begin{table}[t]
	\begin{center}
		\caption{$C_{ij}$ and $Q_{\mathrm{max}}$ for each channels are calculated as below. The uncertainty is dominated by those of measured parameters, especially $a_{\Psi}$. Values for $\eta_c$, $\chi_{c0}$ channels have only trivially small computation uncertainty since they are independent of measured parameters, in contrast to the J/$\psi$ channel. The classical limit for $Q_{\mathrm{max}}$ is $Q_{CL} = 1$.}
			\begin{tabular}{ c | c c c   }
			\multicolumn{4}{c}{} \\			
			\hline
			$$ & $\eta_c$ & $\chi_{c0}$ & J/$\psi$  \\
			\hline
			$C_{11}$ & $\, \, \, \,\, 0.500$  & $-0.500$ & $-0.274 \pm 0.008$ \\
			$C_{22}$ & $-0.500$ & $\, \,  \, \,0.500$  & $-0.177 \pm 0.024$ \\
			$C_{33}$ & $-0.500$ & $-0.500$ & $\,\,\,\,0.404 \pm 0.033$ \\
	       Off-diagonal components & $\,\,\,\,0$	  & $\,\,\,\,\,0$   & $\,\,\,\,0$ 		\\
	       		\hline
		      $Q_{\mathrm{max}}$ & $\,\, 1.414$    & $\,\, 1.414$     & $\,\,\,\,\,\,\,0.976 \pm 0.046$   \\
		  	\hline
		      $Q_{\mathrm{CL}}$ & $\,\,\,\,\,1$    & $$     & $$   \\
			\hline
		\end{tabular}
	\end{center}
	\label{tab.2}
\end{table}
Systematic uncertainties are estimated by shifting the parameters used in the calculation ({\it e.g.} particle masses, $c_{\Lambda}$, $a_{\psi}$ etc.) within their 1$\sigma$ experimental uncertainties. The main contribution is from the uncertainty of $a_{\psi}$ while those from the other parameters have almost no effect on $C$ or $Q_{\mathrm{max}}$. One sees that $Q_{\mathrm{max}}$ in the $\eta_c$ and the $\chi_{c0}$ channel well surpass the classical limit $\mathrm{Q_{CL}} = 1$ and even reach the quantum limit $\sqrt{2}$. In contrast,  the J/$\mathrm{\psi}$ channel gives no significant excess in $Q_{\mathrm{max}}$, therefore, is insensitive to test BI. The $\eta_c$ and the $\chi_{c0}$ channels conserve their entanglement throughout the process whereas the J/$\mathrm{\psi}$ channel loses the spin correlation in J/$\mathrm{\psi \rightarrow \Lambda\overline{\Lambda}}$ where relative orbital angular momentum between the two hyperons dilute the spin correlation by some fraction.  \\
%
 %
\section{Estimation of necessary event number and the corresponding significance}
For the $\mathrm{\eta_c}$ and $\mathrm{\chi_{c0}}$ channels, the BI is violated with the large $Q_{\mathrm{max}}$ values which however experience statistical fluctuations with a limited number of events. In this chapter, we show how the components of $C$ and $Q_{\mathrm{max}}$ fluctuate statistically. Using MC simulations based on matrix (\ref{eq.17}), the distributions of $C_{11}$, $C_{33}$ and $Q_{\mathrm{max}}$ are calculated as Figure 3, where we set the guide axes to the configuration that gives the maximum $Q$ value as we discussed in section 3 i.e. $Q_\mathrm{{max}} = 2 \sqrt{C_{11}^2 + C_{33}^2}$. \\
$C_{11}$ and $C_{33}$ fluctuate according to Gaussian distributions for all $n$. On the other hand, the $Q_{\mathrm{max}}$ distribution has a slight positive bias and can be well approximated by a Gaussian distribution providing $n \small{\gtsim} 1000$. The corresponding mean, RMS and significance are given in Table \ref{tab.3}. The significance is calculated as the corresponding deviation in a Gaussian distribution of the p-value that $Q_{\mathrm{max}}$ fluctuates under the classical limit $Q_{\mathrm{CL}} = 1$. The two channels have the same $n$ dependency and we see about 2000 events are sufficient to announce evidence of the BI violations.   \\ 
\begin{table}[t]
	\caption{Some characteristics of the $Q_{\mathrm{max}}$ distribution}
	\label{tab.3}
	\begin{minipage}[t]{.45\textwidth}
	\begin{center}
			\begin{tabular}{  c | c c c  }
			\multicolumn{4}{c}{} \\			
			\multicolumn{4}{c}{$\eta_c$ channel} \\
			\hline
			$n$ & mean & RMS &significance  \\
			\hline
			$100$ & $1.619$ & $0.653$ & $1.33$ \\
			$300$ & $1.477$ & $0.403$ & $1.55$ \\
			$500$ & $1.454$ & $0.316$ & $1.77$ \\
	                   $1000$ & $1.432$ & $0.223$ & $2.20$ \\
		          $2000$ & $1.423$ & $0.159$ & $2.86$ \\
		          $3000$ & $1.420$ & $0.131$ & $3.43$ \\
		          $5000$ & $1.417$ & $0.100$ & $4.24$ \\ 
			\hline
		\end{tabular}
	\end{center}	
\end{minipage}
\hfill
\begin{minipage}[t]{.45\textwidth}
	\begin{center}
			\begin{tabular}{  c | c c c  }
			\multicolumn{4}{c}{} \\			
			\multicolumn{4}{c}{$\chi_{c0}$ channel} \\
			\hline
			$n$ & mean & RMS &significance  \\
			\hline
			$100$ & $1.617$ & $0.652$ & $1.33$ \\
			$300$ & $1.479$ & $0.405$ & $1.55$ \\
			$500$ & $1.456$ & $0.314$ & $1.77$ \\
	                   $1000$ & $1.433$ & $0.225$ & $2.20$ \\
		          $2000$ & $1.423$ & $0.159$ & $2.86$ \\
		          $3000$ & $1.421$ & $0.131$ & $3.38$ \\
		          $5000$ & $1.418$ & $0.101$ & $4.28$ \\
			\hline
		\end{tabular}
	\end{center}	
	\end{minipage}
\end{table} 
%
%
%
%
%
\begin{figure}[t]
\begin{center}
\label{fig,3}
\includegraphics[width=14.5cm]{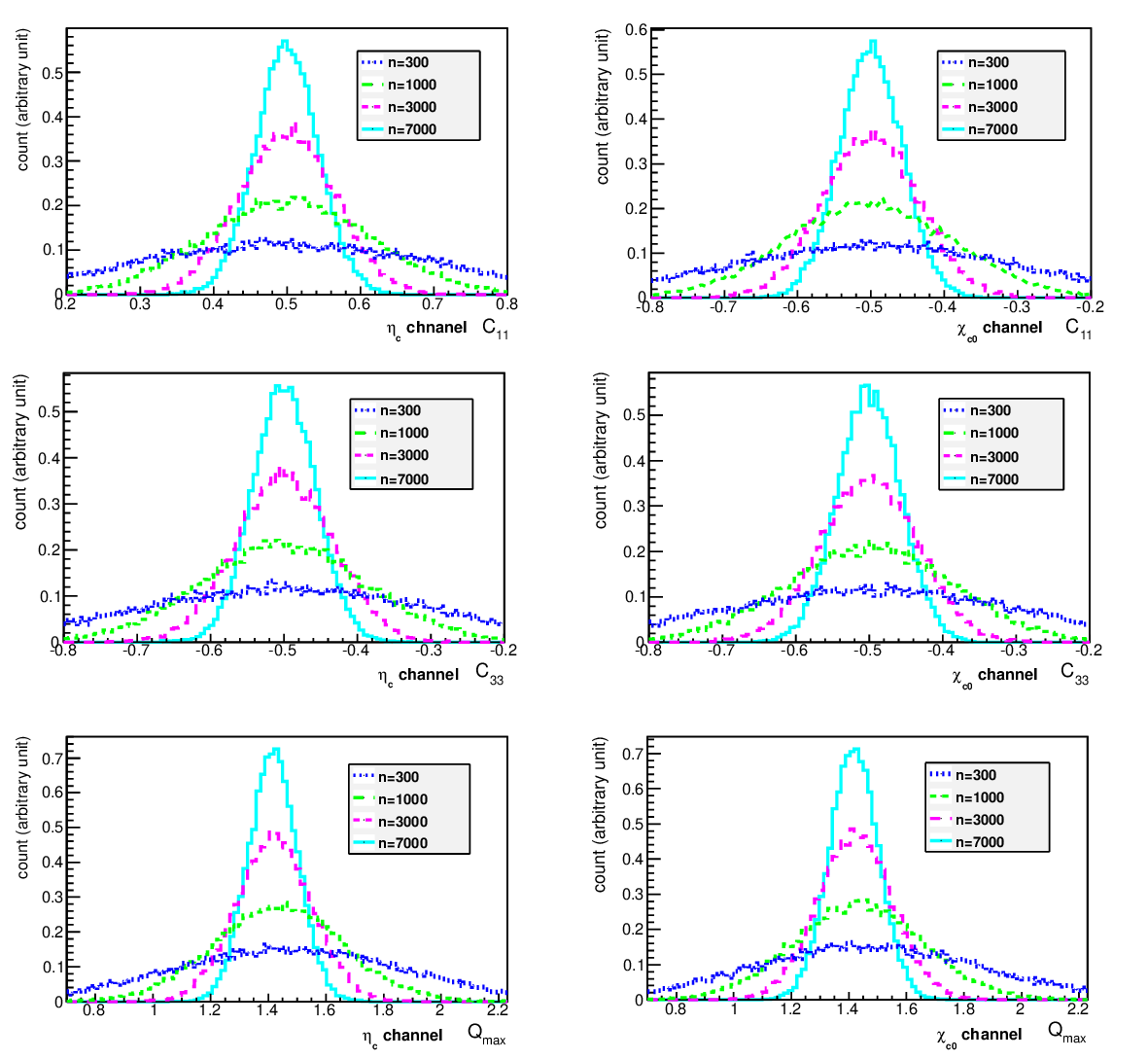}
\caption{Distributions of $C_{11}$, $C_{33}$ and $Q_{\mathrm{max}}$ in the $\eta_c$ and $\chi_{c0}$ channels.}
\end{center}
\end{figure}
%
%
\begin{figure}[H]
\begin{center}
\label{fig,4}
\includegraphics[width=11.0cm]{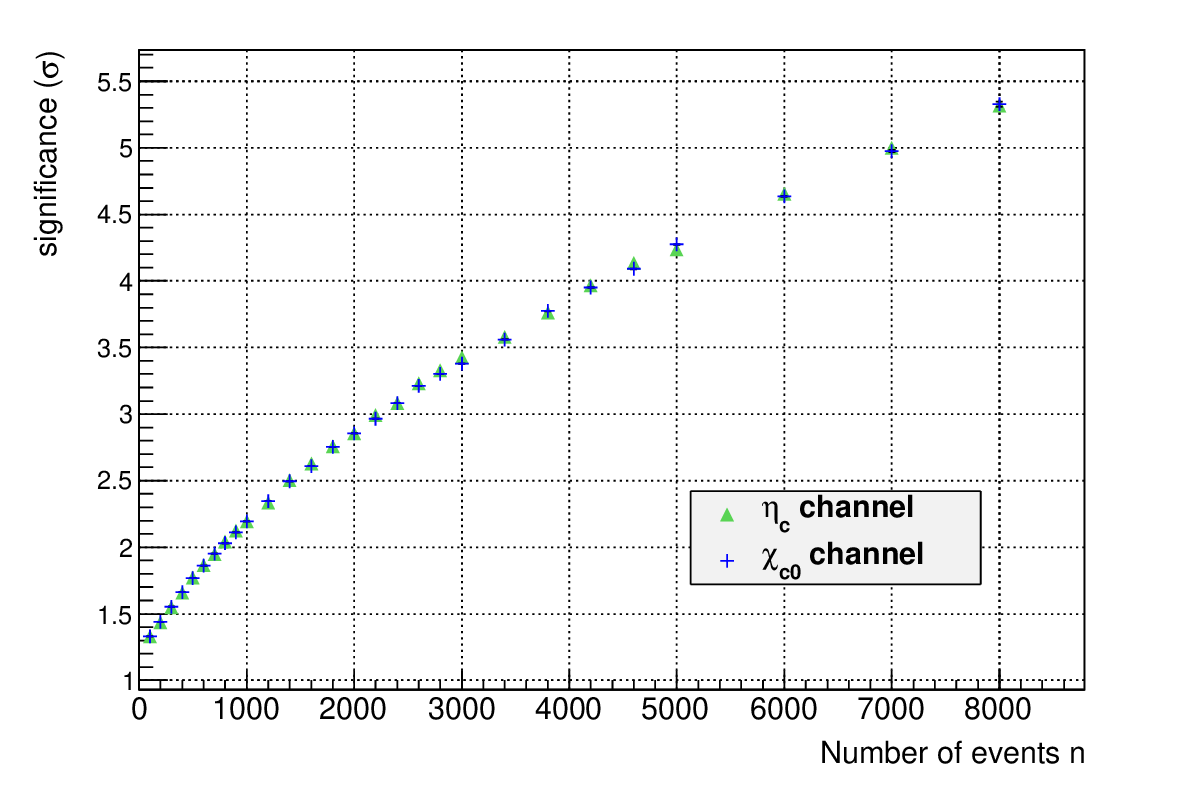}
\caption{Achievable significance with respect to number of events is estimated by MC simulation. We use a large number of samples where MC fluctuation is negligibly small.}
\end{center}
\end{figure}
The number of experimentally available events and the measurement feasibility are also studied. The branching fractions of each relevant decay are listed in Table 4, assuming that $\mathrm{\eta_c}$ and $\mathrm{\chi_{c0}}$ are all produced via $\mathrm{J/\psi  \rightarrow \eta_c + \gamma}$ and $\mathrm{\psi'  \rightarrow \chi_{c0} + \gamma}$ respectively. In the BES3 experiment, $1 \times 10^9$ of J/$\mathrm{\psi}$ and $4 \times 10^8$ of $\psi'$ are planned to be produced by the end of 2012 \cite{BES}. Using only the space-likely separated events which account for around 70$\%$ of all events (Table.1), the event yields are about $6500 \times \epsilon$ and $4000 \times \epsilon$ for the $\eta_c$ and the $\chi_{c0}$ channel respectively, with the event acquisition efficiency $\epsilon$. The $\eta_c$ channel has enough yield even with conservative selection of events (efficiency $\epsilon \sim 0.3$) while the $\chi_{c0}$ channel is available provided a looser selection with $\epsilon \small{\gtsim} 0.6$.
\begin{table}[t]
	\begin{center}
		\caption{Branching fractions of each decay in J/$\mathrm{\psi \rightarrow \gamma \eta_c; \, \eta_c  \rightarrow \Lambda \overline{\Lambda} \rightarrow p{\pi} \overline{p} {\pi}}$ and $\mathrm{\psi' \rightarrow \gamma \chi_{c0}; \, \chi_{c0} \rightarrow \Lambda \overline{\Lambda} \rightarrow p{\pi} \overline{p} {\pi}}$. }
			\begin{tabular}{ c | c  }
			\multicolumn{2}{c}{} \\			
			\hline
			channel & branching fraction \\
			\hline
			$\mathrm{J/\psi  \rightarrow \eta_c + \gamma}$ & $(1.7 \pm 0.4) \times 10^{-2} $  \\
			$\mathrm{\psi'     \rightarrow \chi_{c0} + \gamma}$ & $(9.7 \pm 0.3) \times 10^{-2} $  \\
			$\mathrm{\eta_c      \rightarrow \Lambda + \overline{\Lambda}}$ & $(1.41 \pm 0.17) \times 10^{-3} $  \\
	                   $\mathrm{\chi_{c0} \rightarrow \Lambda + \overline{\Lambda}}$ & $(3.3 \pm 0.4) \times 10^{-4} $  \\
		          $\mathrm{\Lambda           \rightarrow p + \pi^-}$ & $(6.39 \pm 0.05) \times 10^{-1} $  \\
         		          $\mathrm{\overline{\Lambda} \rightarrow \overline{\mathrm{p}} + \pi^+}$ & $(6.39 \pm 0.05) \times 10^{-1}$  \\
			\hline
         	J/$\mathrm{\psi \rightarrow \gamma \eta_c; \, \eta_c \rightarrow \Lambda \overline{\Lambda} \rightarrow p{\pi}p{\pi}}$ & $(9.8 \pm 2.6) \times 10^{-6}$ \\
	$\mathrm{\psi' \rightarrow \gamma \chi_{c0}; \, \chi_{c0} \rightarrow \Lambda \overline{\Lambda} \rightarrow p{\pi}p{\pi}}$ & $(1.31 \pm 0.16) \times 10^{-5}$ \\
			\hline
		\end{tabular}
	\end{center}
	\label{tab.4}
\end{table}
%
%
%
\section{Conclusion}
Charmonium decays $\mathrm{\eta_c \rightarrow \Lambda\overline{\Lambda}}$, $\mathrm{\chi_{c0} \rightarrow \Lambda\overline{\Lambda}}$ and $\mathrm{J/{\psi} \rightarrow \Lambda\overline{\Lambda}}$ are possible probes of Bell's inequality, due to the strongly correlated spins of the hyperon pair. As the hyperons undergo decays in which the angular distribution is associated with the hyperon spins, Bell's inequality can be developed into a new expression in terms of a momentum correlation of the decay products $\mathrm{p}$ and $\mathrm{\overline{p}}$. By QM calculation we find that the correlation in the J/$\mathrm{\psi}$ channel is not strong enough to test Bell's inequality, while it is sufficient in the other two channels. The violation can be experimentally confirmed with around 2000 events for each channel; this number of events would have already been produced at BES, and this experiment is therefore now feasible.\\

\section*{Acknowledgement} 
We would like to thank members and graduates of Komamiya Laboratory (The University of Tokyo) for useful discussions and coorperation, in particular Mr. Daniel Jeans provided outstanding contributions. The authors also appreciate the instruction of  Dr. Koji Hamaguchi (The University of Tokyo), Dr. Izumi Tsutsui (KEK) and the members of the laboratory, whose advice contributed a great deal to this work. This work is also supported by the World Premier International Research Center Initiative (WPI Initiative), MEXT, Japan.\\


\clearpage

\section*{Appendix: The derivation of equation (\ref{eq.4})}
We start with the angle distribution of the protons(anti-protons) from $\Lambda$ ($\overline{\Lambda}$) decays (Same equation as (\ref{eq1}) in the section 1):
\begin{equation}
P(\vec{n} | \vec{s}) =   1 +  \alpha_{\Lambda} \vec{n} \cdot \vec{s}   \mbox{\phantom{MMMMM}}
P (\vec{n'} | \vec{s'}) = 1 -  \alpha_{\bar{\Lambda}} \vec{n'} \cdot \vec{s'} .
\label{eq.A1}
\end{equation}
$\bm{n}(\bm{n}')$ is an unit vector of the the proton (anti-proton) orientation. $P(\bm{n}|\bm{s})$ ($P(\bm{n'}|\bm{s'})$) indicates the conditional probability density of $\vec{n}$ ($\vec{n'}$) with given polarization of the hyperon $\vec{s}$ ($\vec{s'}$). Note that these distributions  are normalized to 1 with $\vec{n}$ ($\vec{n'}$) integration over solid angle $\int d\Omega_{\bm{n}}/4\pi$ ($\int d\Omega_{\bm{n'}}/4\pi$). The correlation amplitude $\braket{(\bm{n}\cdot\bm{a}) (\bm{n}'\cdot\bm{b})}$ can be calculated in terms of these distribution.
\begin{align}
 \braket{(\bm{n}\cdot\bm{a}) (\bm{n}'\cdot\bm{b})} &=\int \frac{d\Omega_{\bm{n}}}{4\pi} \frac{d\Omega_{\bm{n}'}}{4\pi} 
  (\bm{n}\cdot\bm{a}) (\bm{n}'\cdot\bm{b}) \,
  P(\bm{n},\bm{n}') \nonumber \\
  &=\int \frac{d\Omega_{\bm{n}}}{4\pi} \frac{d\Omega_{\bm{n}'}}{4\pi} 
  \frac{d\Omega_{\bm{s}}}{4\pi}
  \frac{d\Omega_{\bm{s'}}}{4\pi}
  (\bm{n}\cdot\bm{a}) (\bm{n}'\cdot\bm{b}) \,
  P
  (\bm{n},\bm{n}' | \bm{s},\bm{s'}) \,
P(\bm{s}, \bm{s'}) \nonumber
\end{align}
$P(x)$ represents the probability density function which a set of variables $x$ follow. When the two hyperon decays are isolated in space-time, which we confirmed in the chapter 2, the joint probability density $P(\bm{n}|\bm{s})P(\bm{n'}|\bm{s'})$ should be identical to the combined probability density $P(\bm{n},\bm{n}' | \bm{s},\bm{s'}) $ since they have no correlation in between, according to the instruction of the locality principle. Using (\ref{eq.A1}), therefore, 
\begin{align}
&\braket{(\bm{n}\cdot\bm{a}) (\bm{n}'\cdot\bm{b})} \nonumber \\
&=\int \frac{d\Omega_{\bm{n}}}{4\pi} \frac{d\Omega_{\bm{n}'}}{4\pi} 
  \frac{d\Omega_{\bm{s}}}{4\pi}
  \frac{d\Omega_{\bm{s'}}}{4\pi}
  (\bm{n}\cdot\bm{a}) (\bm{n}'\cdot\bm{b})
\, P(\bm{n} | \bm{s}) 
\, P(\bm{n'} | \bm{s'}) 
  P(\bm{s},\bm{s'}) \nonumber \\
&=\int \frac{d\Omega_{\bm{n}}}{4\pi} \frac{d\Omega_{\bm{n}'}}{4\pi} 
  \frac{d\Omega_{\bm{s}}}{4\pi}
  \frac{d\Omega_{\bm{s'}}}{4\pi}
  (\bm{n}\cdot\bm{a}) (\bm{n}'\cdot\bm{b})
  (1+\alpha_\Lambda \bm{n}\cdot \bm{s})
  (1-\alpha_{\bar{\Lambda}} \bm{n'}\cdot \bm{s'} )
  P(\bm{s},\bm{s'}). \nonumber
\end{align}
$d\Omega_{\bm{n}} d\Omega_{\bm{n}'}$ integration can be performed using spherical polar coordinates $(\theta,\phi),(\theta',\phi')$ with $\bm{s},\bm{s'}$ being the zenithes:
\begin{align}
	& \braket{(\bm{n}\cdot\bm{a}) (\bm{n}'\cdot\bm{b})} \nonumber \\
  &=\int \frac{d\cos \theta d\phi}{4\pi}
  \frac{d\cos \theta' d\phi'}{4\pi} 
  \frac{d\Omega_{\bm{s}}}{4\pi}
  \frac{d\Omega_{\bm{s'}}}{4\pi}
  (a_1\sin\theta\cos\phi
  +a_2\sin\theta\sin\phi
  +a_3\cos\theta)  \nonumber \\
  &\quad\quad\times(b_1\sin\theta'\cos\phi'
  +b_2\sin\theta'\sin\phi'
  +b_3\cos\theta')
  (1+\alpha_\Lambda \cos \theta)
  (1-\alpha_{\bar{\Lambda}} \cos \theta' )
  P(\bm{s},\bm{s'}) \nonumber
\end{align}
Here $\cos\phi$ and $\sin\phi$ terms vanish in the $\phi$ integral, thus only $\cos\theta$ terms remain. As $a_3=\bm{a}\cdot\bm{s}=s_a$ and $b_3=\bm{b}\cdot\bm{s'}=s'_b$, we obtained the desired form:
\begin{align}
  & \braket{(\bm{n}\cdot\bm{a}) (\bm{n}'\cdot\bm{b})} \nonumber \\
  &=\Bigg(\int \frac{d\cos \theta}{2} 
  \frac{d\cos \theta'}{2}
  (1+\alpha_\Lambda \cos \theta)
  (1-\alpha_{\bar{\Lambda}} \cos \theta' )
  \cos \theta\cos \theta'
  \Bigg) \nonumber \nonumber \\
  &\quad\quad\times\Bigg(\int
  \frac{d\Omega_{\bm{s}}}{4\pi}
  \frac{d\Omega_{\bm{s'}}}{4\pi}
  \, s_a \, s'_b \, P(\bm{s},\bm{s'})
  \Bigg) \nonumber \\
  &=-\frac{\alpha_\Lambda \alpha_{\bar{\Lambda}}}{9}
  \braket{\, s_a \, s'_b \, } \nonumber
\end{align} \\
\end{document}